\documentclass[%
reprint,
 amsmath,amssymb,
 aps,pra,
floatfix,
]{revtex4-2}
\usepackage[T2A,T1]{fontenc}
\usepackage[utf8]{inputenc}
\usepackage{graphicx}
\usepackage{dcolumn}
\usepackage{bm}
\usepackage[colorlinks=true, allcolors=blue]{hyperref}

\usepackage{multirow}
\begin{document}

\title{Diagonal Born--Oppenheimer correction in strong magnetic fields: finite-difference approach for light diatomic molecules}

\author{J. J. Lopez-Rodriguez,$^{1}$ T. Zalialiutdinov,$^{1,2}$ D. Solovyev,$^{1,2}$
}
\affiliation{$^{1}$Department of Physics, St. Petersburg State University, Petrodvorets, Oulianovskaya 1, 198504, St. Petersburg, Russia}
\affiliation{$^{2}$Petersburg Nuclear Physics Institute named after B.P. Konstantinov of National Research Centre 'Kurchatov Institut', St. Petersburg, Gatchina 188300, Russia}

\begin{abstract}
The influence of the diagonal Born–Oppenheimer correction (DBOC) on the electronic structure of light diatomic molecules subjected to strong magnetic fields is investigated. H$_2$, HeH$^+$, and LiH are considered, using both the
Hartree--Fock (HF) approximation and configuration interaction with single and
double excitations (CISD). The analysis covers magnetic field strengths up to
$B_\parallel = 0.2$~a.u.\ ($4.7\times10^{4}$~T), relevant to astrophysical conditions near magnetic white dwarfs.  Although the correction
noticeably shifts total electronic energies, it varies smoothly with internuclear
distance and depends only weakly on the field, so that it enters the vibrational
transition frequencies as a nearly field-independent offset, shifting the
fundamental interval by about $1$~cm$^{-1}$ and the higher levels by up to
$15$~cm$^{-1}$. These shifts exceed the $0.1$--$2.0$~cm$^{-1}$ accuracy currently
attainable for molecular lines in the atmospheres of magnetic white dwarfs.
Comparison of HF and CISD results shows that electron
correlation has a non-negligible effect on the DBOC, most pronounced for the
strongly ionic LiH bond.
\end{abstract}

\maketitle

\section{Introduction}

Molecular systems in strong magnetic fields lie at the interface of fundamental quantum chemistry and high-energy astrophysics. Under the extreme conditions encountered near magnetic white dwarfs and neutron stars, the electronic structure, bonding patterns, and molecular symmetry are substantially modified, rendering these environments qualitatively different from those of conventional chemistry~\cite{10.1063/5.0269984}. At the same time, several light diatomic molecules relevant in these regimes, including HeH$^+$, LiH, and H$_2$, play an important role in astrophysical chemistry \cite{NGC7027, A.Dalgarno_2005, Bovino_2011, LiH_EarlyUniverse}. This motivates the development of accurate theoretical approaches for describing their properties in the presence of strong magnetic fields.

For light few-electron molecules, the standard Born--Oppenheimer (BO) approximation provides a convenient starting point, but it becomes insufficient when the coupling between electronic and nuclear degrees of freedom is non-negligible. The leading-order correction beyond the BO approximation is the diagonal non-adiabatic coupling term (diagonal Born--Oppenheimer correction, DBOC), which accounts for the adiabatic contribution of the nuclear kinetic energy to the electronic energy. Although the DBOC is often small in the absence of external fields, it is by no means negligible and plays an important role in high-precision calculations even at zero field. In the presence of strong magnetic fields, electron--nuclear coupling is further enhanced, which can lead to a substantial increase and nontrivial variation of this contribution \cite{Thorpe17102020}. This effect is particularly pronounced for light molecules with small nuclear masses, for which the DBOC can constitute a significant fraction of the total energy.

In this work, we present a systematic study of the DBOC contribution to the electronic structure of selected light diatomic molecules over a range of strong magnetic field strengths. The electronic structure is treated at the Hartree--Fock (HF) level, with electron correlation included via configuration interaction with single and double excitations (CISD). We compute potential energy curves (PECs) and corresponding vibrational spectra \cite{Vibrationo_H,10.1063/1.4768169_HeH+}, and analyze the evolution of the DBOC contribution as a function of the magnetic field strength.

The results show that the DBOC produces shifts of the vibrational levels of up to about $15$~cm$^{-1}$, with its size relative to the level spacings depending only weakly on the field strength over the range studied, consistent with the trends observed in the potential energy curves and vibrational spectra. These results underscore the importance of including DBOC in quantitative descriptions of molecules under strong-field conditions, with implications for high-energy astrophysics as well as for the emerging field of strong-field molecular chemistry \cite{10.1063/5.0269984}.

The paper is organized as follows. Section~\ref{dboc_theory} presents the
theoretical formulation of the diagonal Born--Oppenheimer correction
for diatomic molecules in a uniform magnetic field and describes its
finite-difference evaluation using phase-corrected wavefunction overlaps.
Computational details are discussed in Section~\ref{computational_details}.
Subsequent sections report potential energy curves and DBOC contributions
for H$_2$, HeH$^{+}$, and LiH at magnetic field strengths
$B=0.0$, $0.10$, $0.15$, and $0.20$~a.u.\ (where
$1~\text{a.u.} = 2.35\times 10^{5}~\text{T}
\approx 2.35\times 10^{9}~\text{G}$). Section~\ref{vibrational}
examines the impact of the DBOC on the vibrational energy levels of these
molecules. Finally, Section~\ref{Conclusion} summarizes the main results and
their implications for precision studies of molecular systems in strong
magnetic fields. Atomic units are used throughout the paper.

\section{Theoretical Background}
\label{dboc_theory}

Within the Born--Oppenheimer approximation, the large mass disparity
between nuclei and electrons allows the electronic wave function to be
treated as adiabatically following the nuclear motion. Then the total
wave function factorizes into an electronic state parametrically dependent
on the nuclear coordinates and a nuclear amplitude evolving on the
resulting potential-energy surface. While this
simplification is widely used and often effective, it becomes quantitatively
inadequate for systems containing light atoms such as hydrogen, where nuclear motion
plays a more pronounced role. Achieving spectroscopic accuracy in such cases requires
explicit corrections beyond the BO framework.

The leading-order correction of this type is the diagonal Born--Oppenheimer correction, defined as the expectation value of the nuclear kinetic energy operator with respect to the electronic wavefunction. For hydrogen-containing molecules, the DBOC is typically larger than both relativistic and quantum electrodynamic contributions, and its inclusion is therefore essential for accurate predictions of vibrational frequencies, dissociation energies, and thermochemical properties~\cite{Valeev2003,Gauss2006}.

We now turn to the specific systems studied in this work, focusing on diatomic molecules in a uniform magnetic field $\mathbf{B}$. Working in the center-of-mass frame with the overall translational motion separated, the internal nuclear coordinate is the relative vector $\mathbf{R}$ between the two nuclei, with conjugate canonical momentum $\hat{\mathbf{P}}$. The nonrelativistic molecular Hamiltonian reads~\cite{Schmelcher1988}
\begin{eqnarray}
\hat{H}_\mathrm{mol}
=
\hat{T}_\mathrm{nuc}
+
\hat{H}_\mathrm{el}(\mathbf{r};\mathbf{R})
+
V_\mathrm{nuc}(\mathbf{R}),
\end{eqnarray}
where $\hat{H}_\mathrm{mol}$ is the total molecular Hamiltonian, $\hat{T}_\mathrm{nuc}$
is the nuclear kinetic-energy operator, $\hat{H}_\mathrm{el}(\mathbf{r};\mathbf{R})$
is the electronic Hamiltonian, depending explicitly on the electronic coordinates
$\mathbf{r}=\{\mathbf{r}_i\}$ and parametrically on the nuclear coordinates
$\mathbf{R}=\{\mathbf{R}_I\}$, and $V_\mathrm{nuc}(\mathbf{R})$ is the Coulomb
repulsion energy between the nuclei.

The nuclear kinetic energy in an external field is
\begin{eqnarray}
\label{nKin}
\hat{T}_\mathrm{nuc}
=
\sum_I \frac{\hat{\boldsymbol{\Pi}}_I^2}{2M_I},
\qquad
\hat{\boldsymbol{\Pi}}_I
=
\hat{\mathbf{P}}_I - Z_I\,\mathbf{A}_\mathrm{ext}(\mathbf{R}_I),
\end{eqnarray}
where $I$ labels the nuclei, $M_I$ and $Z_I$ are the mass and charge number of
nucleus $I$, $\hat{\mathbf{P}}_I$ and $\hat{\boldsymbol{\Pi}}_I$ are its
canonical- and kinetic-momentum operators. The vector potential $\mathbf{A}_\mathrm{ext}(\mathbf{R}_I)=\tfrac{1}{2}\mathbf{B}\times (\mathbf{R}_I-\mathbf{O})$, representing the magnetic contribution to the nuclear kinetic momentum, is evaluated at $\mathbf{R}_I$, with $\mathbf{O}$ being the (arbitrary) gauge origin. The electronic
Hamiltonian defines the Born--Oppenheimer surface through
\begin{eqnarray}
\hat{H}_\mathrm{el}(\mathbf{r};\mathbf{R})\,
|\Psi(\mathbf{R})\rangle
=
U_\mathrm{BO}(\mathbf{R})\,
|\Psi(\mathbf{R})\rangle,
\end{eqnarray}
where $|\Psi(\mathbf{R})\rangle$ is the electronic wave function at fixed
$\mathbf{R}$ and $U_\mathrm{BO}(\mathbf{R})$ is the corresponding potential-energy
surface.

Within the Born–Oppenheimer approximation, the field-dependent electronic Hamiltonian $\hat{H}_\mathrm{el}$ of an $N$-electron molecule can be written as \cite{10.1063/5.0269984}
\begin{eqnarray}
\label{1}
    \hat{H}_\mathrm{el} = \hat{H}_0 + \frac{1}{2} \sum_{i=1}^{N} \mathbf{B} \cdot \mathbf{l}_i^{\mathrm{O}} + \mathbf{B} \cdot \mathbf{S}
    \\\nonumber
    + \frac{1}{8} \sum_{i=1}^{N} \left[ B^2 (r_i^{\mathrm{O}})^2 - (\mathbf{B} \cdot \mathbf{r}_i^{\mathrm{O}})^2 \right].
\end{eqnarray}
Here $\hat{H}_0$ is the usual field-free electronic Hamiltonian (kinetic energy
plus electron-electron and electron-nucleus Coulomb terms, for $N$ electrons), $B=|\mathbf{B}|$, $\mathbf{r}_i^{\mathrm{O}}=\mathbf{r}_i-\mathbf{O}$
is the position of electron $i$ relative to the gauge origin $\mathbf{O}$,
$r_i^{\mathrm{O}}=|\mathbf{r}_i^{\mathrm{O}}|$,
$\mathbf{l}_i^{\mathrm{O}}=-\mathrm{i}[\mathbf{r}_i^{\mathrm{O}}\times\nabla_i]$
is the orbital angular-momentum operator of electron $i$ relative to
$\mathbf{O}$, and $\mathbf{S}=\sum_{i=1}^N\mathbf{s}_i$ is the total electron-spin
operator, with $\mathbf{s}_i$ the spin operator of electron $i$
\cite{Stopkowicz_2015}. The conventional description of the additional contributions in Eq.~(\ref{1}) is as follows. The second term is the orbital paramagnetic interaction, the third is the spin paramagnetic (Zeeman) interaction, and the fourth is the diamagnetic interaction.

The explicit dependence of Eq.~(\ref{1}) on $\mathbf{r}_i^{\mathrm{O}}$ reflects
the gauge dependence of the vector potential: a change of gauge origin,
$\mathbf{O}\to\mathbf{G}$, shifts $\mathbf{A}$ by the gradient of a scalar
function, $\mathbf{A}_{\mathbf G}(\mathbf{r}_i)=\mathbf{A}_{\mathbf
O}(\mathbf{r}_i)+\nabla f$. As a result, the exact wave function should acquire a compensating phase
\begin{eqnarray}
    \label{psi}
    \Psi_\mathrm{G} =
    \exp{\left[\frac{\mathrm{i}}{2}\mathbf{B}\times (\mathbf{G}-\mathbf{O}) \cdot\mathbf{r} \right]} \Psi_\mathrm{O},
\end{eqnarray}
where $\Psi_\mathrm{O}$ and $\Psi_\mathrm{G}$ are the wave functions referred to
gauge origins $\mathbf{O}$ and $\mathbf{G}$, respectively. The transformation (\ref{psi}) leaves physical observables (evaluated as expectation values of Hermitian operators) invariant under this phase
\cite{Landau:1991wop}.

By incorporating this phase into the atomic-orbital basis, one obtains a basis function centered at $\mathbf{K}$ that depends explicitly on the field and the gauge, given by
\begin{eqnarray}
    \label{ao}
    \omega_{lm}(\mathbf{r}_{\mathrm{K}}, \mathbf{B}, \mathbf{G}) = \exp \left[ \frac{ \mathrm{i}}{2} \mathbf{B} \times (\mathbf{G} - \mathbf{K}) \cdot \mathbf{r} \right]
    g_{lm}(\mathbf{r}_{\mathrm{K}}).\,\,\,
\end{eqnarray}
Here $\mathbf{K}$ is the center (nucleus) of the basis function,
$g_{lm}(\mathbf{r}_{\mathrm{K}})$ is the ordinary field-free atomic orbital
with angular-momentum quantum numbers $l,m$ centered at $\mathbf{K}$, and
$\omega_{lm}(\mathbf{r}_{\mathrm{K}},\mathbf{B},\mathbf{G})$ is the resulting
field-dependent orbital. This construction, first proposed by Fritz London
\cite{London1937}, is known as the London orbital, or GIAO (gauge-including,
equivalently gauge-origin-independent, atomic orbital).

In summary, the Hamiltonian~(\ref{1}) is gauge-origin dependent and contains
the electronic orbital angular-momentum operator, so its eigenfunctions are in
general complex. The use of GIAOs, Eq.~(\ref{ao}), correspondingly modifies the
one- and two-electron integrals of standard quantum-chemistry codes \cite{Irons2017}. In strong fields, the Born--Oppenheimer approximation itself requires modification through nonadiabatic coupling terms (both diagonal and, near degeneracies, off-diagonal), yet it remains a practical and widely used starting point.

Within the BO approximation the total wavefunction factorizes
as $|\Phi\rangle = |\Psi(\mathbf{R})\rangle|\chi(\mathbf{R})\rangle\equiv |\Psi(\mathbf{R})\rangle\,\chi$, and projecting the full
Schr\"odinger equation onto $|\Psi(\mathbf{R})\rangle$ gives the effective
nuclear equation:
\begin{eqnarray}
&{}&
\hat{H}_\mathrm{nuc}\,\chi = E\,\chi,
\label{eq:Hnuc_def}
\\
\nonumber
&{}&
\hat{H}_\mathrm{nuc}
=
\langle\Psi|\hat{T}_\mathrm{nuc}|\Psi\rangle
+
U_\mathrm{BO}
+
V_\mathrm{nuc}.
\end{eqnarray}
In this equation, $|\Psi(\mathbf{R})\rangle$ is the electronic wavefunction that depends parametrically on $\mathbf{R}$, $\chi(\mathbf{R})$ is the nuclear wavefunction governing the vibrational and rotational motion on the Born--Oppenheimer surface $U_\mathrm{BO}(\mathbf{R})$, and $E$ is the total energy eigenvalue.

The nuclear kinetic energy operator $\hat{T}_\mathrm{nuc}$ is applied to $|\Phi\rangle$ separately for each nucleus, as indicated in Eq.~(\ref{nKin}). Dropping the index $I$ for brevity, we consider a single term $\hat{\boldsymbol{\Pi}}^2/(2M)$. Since $\mathbf{A}_\mathrm{ext}$ depends solely on the nuclear coordinates, the resulting expression is readily evaluated.
\begin{eqnarray}
\hat{\boldsymbol{\Pi}}\,\bigl[|\Psi\rangle\chi\bigr]
=
|\Psi\rangle\,\hat{\boldsymbol{\Pi}}\chi
+\bigl(\hat{\mathbf{P}}|\Psi\rangle\bigr)\chi.
\end{eqnarray}
By applying $\hat{\boldsymbol{\Pi}}$ a second time and projecting with $\langle\Psi|$, we find
\begin{align}
\langle\Psi|\hat{\boldsymbol{\Pi}}^{2}|\Psi\rangle\chi
=
\hat{\boldsymbol{\Pi}}^{2}\chi
+
2\,\boldsymbol{{\mathcal A}}
  \cdot\hat{\boldsymbol{\Pi}}\chi
+
\langle\Psi|\hat{\mathbf{P}}^2|\Psi\rangle\,\chi,
\label{eq:projected_Pi2}
\end{align}
where $\boldsymbol{{\mathcal A}}$ is the Berry connection (geometric vector
potential \cite{Berry1984,MeadTruhlar1979})
\begin{eqnarray}
\boldsymbol{{\mathcal A}}(\mathbf{R})
\equiv
\langle\Psi(\mathbf{R})|
  \hat{\mathbf{P}}
|\Psi(\mathbf{R})\rangle
= -i\langle\Psi|\nabla\Psi\rangle.
\label{eq:Berry_connection}
\end{eqnarray}

It is worth noting that differentiating the normalization condition $\langle\Psi|\Psi\rangle=1$ with respect to $R_i$ gives $\langle\partial_{R_i}\Psi|\Psi\rangle+\langle\Psi|\partial_{R_i}\Psi\rangle=0$. Since the two terms are conjugates of each other, $\langle\partial_{R_i}\Psi|\Psi\rangle = \langle\Psi|\partial_{R_i}\Psi\rangle^{*}$, one obtains $2\,\mathrm{Re}\,\langle\Psi|\partial_{R_i}\Psi\rangle=0$, which means that $\langle\Psi|\partial_{R_i}\Psi\rangle$ is purely imaginary. Consequently, ${\mathcal A}_{i}=-\mathrm{i}\langle\Psi|\partial_{R_i}\Psi\rangle$ is real-valued for any normalized wavefunction, real or complex.



In the absence of a magnetic field, both the electronic Hamiltonian and the wavefunction $|\Psi\rangle$ can be taken as real. In that case $\langle\Psi|\partial_{R_i}\Psi\rangle$ is also real, and as shown above $\boldsymbol{{\mathcal A}}=0$ identically. In contrast, in the presence of a magnetic field, the orbital paramagnetic term renders the Hamiltonian $\hat{H}_\mathrm{el}$ intrinsically complex, see Eq.~(\ref{1}). As a consequence, due to the complex nature of the wave function, the scalar product $\langle\Psi|\partial_{R_i}\Psi\rangle$ reduces to its imaginary part, and $\boldsymbol{\mathcal A}$ is generally nonzero. Moreover, the Berry connection becomes gauge-dependent under a geometry-dependent phase transformation $|\Psi\rangle\to e^{-\mathrm{i}F(\mathbf{R})}|\Psi\rangle$. For real $F$, it transforms as $\boldsymbol{\mathcal A}\to\boldsymbol{\mathcal A}-\nabla F$, whereas all physical observables are left invariant.

Subsequently, by adding and subtracting $\boldsymbol{{\mathcal A}}^2\chi$ to complete the square
\begin{eqnarray}
\hat{\boldsymbol{\Pi}}^{2}\chi
+2\,\boldsymbol{{\mathcal A}}\cdot\hat{\boldsymbol{\Pi}}\chi
+\boldsymbol{{\mathcal A}}^2\chi
=
\bigl(\hat{\boldsymbol{\Pi}}+\boldsymbol{{\mathcal A}}\bigr)^2\chi
\equiv
\tilde{\boldsymbol{\Pi}}^{2}\chi,
\label{eq:square}
\end{eqnarray}
one can define the screened nuclear momentum as \cite{Culpitt2021}
\begin{eqnarray}
\tilde{\boldsymbol{\Pi}}
\equiv
\hat{\mathbf{P}}
- Z\,\mathbf{A}_\mathrm{ext}(\mathbf{R})
+\boldsymbol{{\mathcal A}}(\mathbf{R}).
\label{eq:screened_momentum}
\end{eqnarray}
Back-substitution of Eq.~\eqref{eq:square} into Eq.~\eqref{eq:projected_Pi2} results in
\begin{eqnarray}
\langle\Psi|\hat{\boldsymbol{\Pi}}^{2}|\Psi\rangle\chi
=
\tilde{\boldsymbol{\Pi}}^{2}\chi
+
\Bigl[
  \langle\Psi|\hat{\mathbf{P}}^{2}|\Psi\rangle
  -\boldsymbol{{\mathcal A}}^2
\Bigr]\chi.
\label{eq:Pi2_split}
\end{eqnarray}

Using the completeness condition for the wavefunctions $\Psi$ in the matrix element of the bracketed expression in Eq.~(\ref{eq:Pi2_split}), one finds that
\begin{eqnarray}
\langle\Psi|\hat{\mathbf{P}}^{2}|\Psi\rangle
-\boldsymbol{{\mathcal A}}^2
=
\sum_{p\neq 0}
|\langle\Psi_p|\hat{\mathbf{P}}|\Psi\rangle|^2
\equiv
-\,\langle\Psi|\Delta'|\Psi\rangle,
\label{eq:Delta_def}
\end{eqnarray}
where the $p=0$ term exactly cancels $\boldsymbol{{\mathcal A}}^2$ (see Eq.~(\ref{eq:Berry_connection})).
Then, upon restoring the nuclear index $I$, the effective nuclear Hamiltonian takes the form~\cite{Culpitt2021}:
\begin{eqnarray}
\hat{H}_\mathrm{nuc}
=
\sum_I \frac{\tilde{\boldsymbol{\Pi}}_I^{2}}{2M_I}
+
U_\mathrm{BO}(\mathbf{R})
+
V_\mathrm{nuc}(\mathbf{R})
+
U_\mathrm{DBOC}(\mathbf{R}),\,\,\,
\label{eq:Hnuc_final}
\end{eqnarray}
with the diagonal Born--Oppenheimer correction
\begin{eqnarray}
U_\mathrm{DBOC}(\mathbf{R})
=-
\sum_I \frac{1}{2M_I}\,\langle\Psi |\Delta'_I|\Psi\rangle.
\label{eq:DBOC}
\end{eqnarray}
The prime on the Laplace operator in definition (\ref{eq:Delta_def}) indicates the omission of the zero-momentum contribution. In the absence of a magnetic field ($\mathbf{B}=0$), $\boldsymbol{\mathcal A}=0$, and $\Delta'_I$ reduces to the ordinary Laplacian $\Delta_I=\sum_i\partial^2/\partial R_{I,i}^2$.

Direct computation of $\Delta'_I$ In Eq. (\ref{eq:Delta_def}) would require coupled-perturbed equations for the analytical wavefunction derivatives. A more practical route corresponds to the finite-difference overlap scheme \cite{Valeev2003}, which can be applied to any electronic-structure method without explicit differentiation of the wavefunction. Within this approach, the position of nucleus $I$ is displaced by $\pm\delta r_{i}$ along each Cartesian direction $i\in\{x,y,z\}$, and the overlap integrals
\begin{eqnarray}
    S^{(\pm)}_{I,i} = \langle \Psi(\mathbf{R}_0) |
    \Psi(\mathbf{R}_0 \pm \delta\mathbf{r}_i) \rangle
\end{eqnarray}
allow the second derivative to be approximated by the central-difference
formula~\cite{Valeev2003}:
\begin{eqnarray}
    \left\langle \Psi \middle| \frac{\partial^2}{\partial R_{I,i}^2}
    \middle| \Psi \right\rangle_{\!\mathbf{R}_0}
    \approx \frac{S^{(+)}_{I,i} + S^{(-)}_{I,i} - 2}{\delta r_{i}^2}.
    \label{eq:laplacian_overlap}
\end{eqnarray}

Summing over Cartesian directions, the Laplacian expectation value reads
\begin{eqnarray}
    \left\langle \Psi \middle| \Delta_{I} \middle| \Psi \right\rangle
    \approx \sum_{i=x,y,z}
    \frac{S^{(+)}_{I,i} + S^{(-)}_{I,i} - 2}{\delta r_{i}^2}.
    \label{eq:laplacian_components}
\end{eqnarray}
In the Hartree--Fock case, $\Psi$ is a Slater determinant and the overlap
factorizes into spin components,
\begin{eqnarray}
    S^{(\pm)}_{I,i} = S_\alpha^{(\pm)}\, S_\beta^{(\pm)}, \qquad
    S_\sigma^{(\pm)} = \det\!\left[
        \mathbf{C}^{\sigma\dagger}_0\, \mathbf{S}^{(\pm)}\, \mathbf{C}^\sigma_\pm
    \right],
    \label{eq:HF_overlap_det}
\end{eqnarray}
where $\mathbf{C}^\sigma_0$, $\mathbf{C}^\sigma_\pm$ are the MO
coefficient matrices at the reference and displaced geometries, respectively, and
$(\mathbf{S}^{(\pm)})_{pq} = \langle\phi_p(\mathbf{R}_0)|
\phi_q(\mathbf{R}_0\pm\delta\mathbf{r}_i)\rangle$
is the AO overlap matrix. For wavefunctions determined within the restricted Hartree--Fock approach, $S^{(\pm)}_{I,i}=\bigl(S_\alpha^{(\pm)}\bigr)^2$.

In a strong magnetic field, however, the SCF cycle at each displaced
geometry can converge to the correct occupied subspace but with an essentially
arbitrary overall phase and unitary rotation of the occupied orbitals.
Writing each overlap in polar form,
\begin{equation}
    S^{(\pm)}_{I,i} = \big|S^{(\pm)}_{I,i}\big|\, e^{\mathrm{i}\varphi^{(\pm)}_{I,i}},
    \qquad
    \varphi^{(\pm)}_{I,i} = \arg S^{(\pm)}_{I,i},
    \label{eq:polar}
\end{equation}
reveals the degree of freedom. The SCF phase noise that affects Eq.~\eqref{eq:laplacian_components} arises from the rapid variation of $\varphi^{(\pm)}_{I,i}$ between displaced geometries, even when $|S^{(\pm)}_{I,i}|\approx 1$ and the electronic state changes only slightly~\cite{Culpitt2021}.

The remedy is to phase-correct each displaced overlap by canceling its argument:
\begin{equation}
    S^{(\pm)}_{I,i} \;\longrightarrow\;
    \widetilde{S}^{(\pm)}_{I,i} = e^{-\mathrm{i}\varphi^{(\pm)}_{I,i}}\, S^{(\pm)}_{I,i}
    = \big|S^{(\pm)}_{I,i}\big|.
    \label{eq:rephasing}
\end{equation}
This discards the arbitrary phase while retaining the physical amplitude.
Inserting the rephased overlaps into Eq.~\eqref{eq:laplacian_components} gives the
phase-corrected expression:
\begin{equation}
    \langle\Psi|\Delta'_{I}|\Psi\rangle
    \;\approx\; \sum_{i=x,y,z}
    \frac{|S^{(+)}_{I,i}| + |S^{(-)}_{I,i}| - 2}{\delta r_{i}^{2}}.
    \label{eq:laplacian_modulus}
\end{equation}
It is this quantity that enters Eq.~\eqref{eq:DBOC}, thus avoiding any explicit evaluation of the Berry connection.

To demonstrate that the Berry connection can be disregarded, one may write
$\langle\Psi|\partial_{I,i}\Psi\rangle = \mathrm{i}\,{\mathcal A}_{I,i}$ with
${\mathcal A}_{I,i}$ real. Then, expanding the displaced wavefunction to second order in the displacement, we find
\begin{equation}
    S^{(\pm)}_{I,i}
    = 1 \pm \mathrm{i}\,{\mathcal A}_{I,i}\,\delta r_i
      + \tfrac{1}{2}\,\delta r_i^{2}\,
        \langle\Psi|\partial_{I,i}^{2}\Psi\rangle
      + O(\delta r_i^{3}) .
    \label{eq:overlap_expansion}
\end{equation}
The direction of the displacement enters only through the linear term, which is purely
imaginary and therefore contributes to the modulus at second order. Using
$\mathrm{Re}\,\langle\Psi|\partial_{I,i}^{2}\Psi\rangle
=-\langle\partial_{I,i}\Psi|\partial_{I,i}\Psi\rangle$, which follows from
differentiating the normalization condition twice, we obtain
\begin{equation}
    \big|S^{(\pm)}_{I,i}\big|
    = 1-\frac{\delta r_i^{2}}{2}     \Bigl(\langle\partial_{I,i}\Psi|\partial_{I,i}\Psi\rangle
            -{\mathcal A}_{I,i}^{2}\Bigr)
      +O(\delta r_i^{4}).
    \label{eq:modulus_expansion}
\end{equation}
The expression (\ref{eq:modulus_expansion}) is the same for both displacements. Adding the two moduli therefore doubles the quadratic term, so that, up to $O(\delta r_i^{2})$, the right-hand side of Eq.~\eqref{eq:laplacian_modulus} simplifies to
\begin{eqnarray}
&{}& \frac{|S^{(+)}_{I,i}|+|S^{(-)}_{I,i}|-2}{\delta r_i^{2}}
\nonumber\\
&{}& \quad = -\Bigl(\langle\partial_{I,i}\Psi|\partial_{I,i}\Psi\rangle
        -\bigl|\langle\Psi|\partial_{I,i}\Psi\rangle\bigr|^{2}\Bigr)
        +O(\delta r_i^{2})
\nonumber\\
&{}& \quad = -\sum_{p\neq 0}
        \bigl|\langle\Psi_p|\partial_{I,i}\Psi\rangle\bigr|^{2}
    = \langle\Psi|\Delta'_{I,i}|\Psi\rangle .
\label{eq:metric_identity}
\end{eqnarray}
Here the last step follows from the definition~\eqref{eq:Delta_def}. Therefore, Eq.~\eqref{eq:laplacian_modulus} is verified to the order of the finite-difference approximation.

Another important implication of Eq.~(\ref{eq:metric_identity}) is that the Berry connection is automatically removed upon substitution of the overlap moduli in the full Laplacian operator, $\langle\Psi|\Delta'_{I,i}|\Psi\rangle =\langle\Psi|\Delta_{I,i}|\Psi\rangle+{\mathcal A}_{I,i}^{2}$. Thus, Eqs.~\eqref{eq:DBOC} and \eqref{eq:laplacian_modulus} imply that a single set of re-phased overlap matrices yields the DBOC directly, with no need for coupled-perturbed equations or a separate computation of $\boldsymbol{\mathcal A}$. Furthermore, this rephasing corresponds to a local gauge transformation, $|\Psi\rangle\to e^{-\mathrm{i}F(\mathbf{R})}|\Psi\rangle$ and $\boldsymbol{\mathcal{A}}\to\boldsymbol{\mathcal{A}}-\nabla F$. Hence, with a suitable choice of the reference geometry, the Berry connection $\boldsymbol{\mathcal{A}}$ can be set to zero identically. In this context, Eq.~(\ref{eq:metric_identity}) is gauge invariant owing to the subtraction of the two terms inside the parentheses. Since no assumption regarding the reality of the wavefunctions has been introduced, Eq.~\eqref{eq:DBOC} recovers the standard DBOC expression in the absence of an external field.


Finally, a noteworthy point is that the overlap matrices do not depend on the gauge origin of the vector potential. A change of origin $\mathbf{O}\to\mathbf{G}$ multiplies the electronic wavefunction at every nuclear geometry by the same many-electron phase factor, Eq.~\eqref{psi}, which cancels between bra and ket vectors in $S^{(\pm)}_{I,i}$. The expression for the DBOC is therefore inherently independent of the gauge origin.

In summary, the developed framework demonstrates that treating strong, non-perturbative external magnetic fields requires two independent phase adaptations to maintain gauge invariance. First, the gauge dependence of the orbital-, spin-paramagnetic, and diamagnetic contributions in the modified interaction Hamiltonian, Eq.~\eqref{1}, which shifts all required calculations into the complex domain, is countered by introducing London phase factors into the atomic-orbital basis, see~\cite{London1937} and Eq.~\eqref{ao}. Second, the presence of a magnetic field modifies the Born--Oppenheimer approximation itself, where the nuclear Hamiltonian yields the Berry connection, Eq.~\eqref{eq:projected_Pi2}, and a screened nuclear momentum, Eq.~\eqref{eq:screened_momentum}. In a finite-difference approach together with rephasing, the resulting geometric phases in the complex matrix elements are shown to exactly cancel the Berry connection in the diagonal Born--Oppenheimer correction, Eq.~\eqref{eq:DBOC}. By replacing these complex overlap matrix elements with their moduli according to Eqs.~\eqref{eq:laplacian_modulus} and~\eqref{eq:metric_identity}, the final DBOC remains rigorously not only independent of both phases but also gauge-invariant.


For a multi-determinant CISD wavefunction, the overlap $S^{(\pm)}_{I,i}$ cannot be reduced to a single determinant expression. Instead, it is evaluated as the full overlap between two CISD expansions at different nuclear geometries, followed by the same phase correction, $\widetilde{S}^{(\pm)}_{I,i} = |S^{(\pm)}_{I,i}|$, before substitution into Eq.~\eqref{eq:laplacian_components}. These overlaps are evaluated using routines implemented in \texttt{PySCF}~\cite{PySCF}. In the unrestricted case, the $\alpha$ and $\beta$ spin channels are treated independently with separate MO coefficient matrices $\mathbf{C}^{\alpha}$ and $\mathbf{C}^{\beta}$, and the resulting overlaps are substituted into Eq.~\eqref{eq:laplacian_components} exactly as in the single-determinant case.

\section{Computational Details}
\label{computational_details}

We perform all electronic structure calculations using the uncontracted correlation-consistent polarized valence triple-zeta (unc-cc-pVTZ) basis set. It provides a balanced description of the electronic wave functions for the light molecular systems considered in this work. The ground-state potential energy curves were computed at the Hartree--Fock (HF) level, with electron correlation included via configuration interaction with single and double excitations (CISD). 

The magnetic field was applied parallel to the molecular axis ($\mathbf{B} \parallel \hat{z}$), and the field-dependent one- and two-electron integrals were evaluated using gauge-including atomic orbitals (GIAOs, or London orbitals) to ensure gauge-origin independence of the results. The modified integrals were generated using the \texttt{ChronusQ} quantum chemistry package~\cite{ChronusQ} and subsequently passed to the \texttt{PySCF} program library~\cite{PySCF} for the correlated calculations.

The diagonal Born--Oppenheimer correction was evaluated using the finite-difference overlap scheme described in Section~\ref{dboc_theory}. For each nucleus $I$, its position was shifted by $\delta r = 10^{-4}$~a.u. along each of the three Cartesian directions ($x$, $y$, $z$). Then, the overlap integrals between the reference and displaced geometries were computed. The step size $\delta r = 10^{-4}$~a.u.\ was chosen to balance the truncation error of the central-difference approximation against numerical noise arising from the SCF convergence and the finite precision of the integral evaluation. Tests with step sizes in the range $\delta r = 10^{-3}$--$10^{-5}$~a.u.\ confirmed that the DBOC values are converged to within $\sim 10^{-7}$~a.u.\ at $\delta r = 10^{-4}$~a.u. The phase correction of Eq.~\eqref{eq:rephasing}, described in Section~\ref{dboc_theory}, was applied to remove gauge noise from the complex-valued overlaps in the presence of the magnetic field, ensuring that the final DBOC values are numerically stable and gauge-invariant.

In the present implementation, the cross-geometry atomic-orbital overlaps used for
both the UHF and UCISD wavefunctions were evaluated with the field-free functions,
i.e.\ without the London phase factors. For the parallel orientation these phases
vanish at the reference geometry and for nuclear shifts along the molecular axis,
but not for those perpendicular to the field. Our estimates indicate that this
approximation adds a small, nearly $R$-independent contribution to the DBOC, which
largely cancels in the vibrational transition frequencies.

To validate the finite-difference approach, we performed benchmark calculations at zero magnetic field ($B = 0$) using the \texttt{CFOUR} quantum chemistry package~\cite{CFOUR}, which implements the DBOC via analytical second derivatives of the wave function at the CCSD level. The comparison between our finite-difference CISD results and the analytical CCSD values from \texttt{CFOUR} shows agreement, with deviations typically on the order of $10^{-5}$--$10^{-6}$~a.u. for the equilibrium geometries of H$_2$, HeH$^{+}$, and LiH. This level of agreement confirms the accuracy and reliability of the finite-difference overlap scheme employed in this work. The result is especially noteworthy since the method had to be adapted to complex-valued wave functions in strong magnetic fields, where analytical derivatives are not currently available.

The limit of large interatomic distance $R$ serves as a second, parameter-free test. For large internuclear separations, the electronic wavefunction of each fragment depends on the nuclear position solely through a rigid translation of all its electrons. Consequently, by virtue of the momentum conservation law for an isolated atom,
$\partial_{\mathbf{R}_A}\Psi=-\sum_{i\in A}\nabla_i\Psi$ and
\begin{eqnarray}
    U_\mathrm{DBOC}\;\longrightarrow\;\sum_A\frac{1}{2M_A}
    \Bigl\langle\Bigl(\textstyle\sum_{i\in A}\hat{\mathbf p}_i\Bigr)^{2}\Bigr\rangle
    \\\nonumber
    =\sum_A\frac{1}{2M_A}\Bigl[2\langle T_A\rangle
    +2\!\!\sum_{i<j\in A}\!\langle\hat{\mathbf p}_i\cdot\hat{\mathbf p}_j\rangle\Bigr].
\end{eqnarray}
Here the first term is fixed by the electronic kinetic energy of the isolated atom and the second is the mass-polarization (specific mass shift) operator~\cite{HughesEckart1930,DrakeBook}. The reduced mass arising in the above expression is, to a sufficiently good approximation, replaced by the atomic mass. For a one-electron fragment, this term is inherently absent, and the virial theorem yields $\langle T_A\rangle=-E_A$, meaning that the DBOC limit follows from the atomic energy alone. This is the case for H$_2$: with $E_A=E_{\rm H}=-0.499826$~a.u., taken as one half of the computed H$_2$ energy at $R=10$~bohr, and $M_A=M_p=1836.15\,m_e$, the predicted limit is $119.5$~cm$^{-1}$. The computed curve gives $119.4$~cm$^{-1}$ at $R=10$~bohr, in agreement to within $0.1\%$.

For HeH$^{+}$ only the helium atom contributes, since the proton carries no electrons of its
own, but the mass-polarization term no longer vanishes. Neglecting it would predict
$87.4$~cm$^{-1}$, whereas including the cross term with the accurate helium value
$\langle\hat{\mathbf p}_1\cdot\hat{\mathbf p}_2\rangle=0.15907$~a.u. ~\cite{Pekeris1959,DrakeBook} gives $92.2$~cm$^{-1}$, in agreement with the computed plateau of
$92.0$~cm$^{-1}$. Conversely, starting from the DBOC curve at infinite internuclear separation, we obtain $\langle\hat{\mathbf p}_1\cdot\hat{\mathbf p}_2\rangle = 0.152$~a.u., about $4\%$ lower than the exact value. This deviation is not unexpected, since the mass-polarization expectation value is more sensitive to electron-correlation effects than the total energy, and such discrepancies are typical at the CISD level with a triple-zeta basis. For LiH the same analysis is less conclusive, since the molecule is still bound by about
$18$~mE$_{\rm h}$ at $R=10$~bohr and has not yet reached the dissociation limit. Finally, the DBOC computed via the proposed approach reproduces not only the atomic kinetic energies but also the mass-polarization contribution, ensuring that both terms enter the dissociation limit on an equal footing.

To evaluate the impact of the diagonal Born--Oppenheimer correction on the vibrational spectrum, the corresponding energy levels were determined by solving the one-dimensional nuclear Schr\"odinger equation. This was performed on the computed potential energy curves using the \texttt{Duo} spectroscopic modeling software~\cite{Duo}. The nuclear motion calculations were restricted to a field-free radial equation solved explicitly for the $J=0$ rotational state. With the gauge origin placed at the center of mass and $\mathbf{B}\parallel\hat{z}$, the vector potential acting on the relative nuclear coordinate, $\mathbf{A}_\mathrm{ext}(\mathbf{R})=\tfrac{1}{2}\mathbf{B}\times\mathbf{R}$, vanishes when $\mathbf{R}$ is parallel to $\mathbf{B}$.
Consequently, in the parallel geometry considered here, the magnetic field enters the vibrational problem only implicitly, through the electronic potential energy curve. These potential energy curves are interpolated within the \texttt{Duo} package using cubic splines, with the vibrational eigenvalues being determined by diagonalizing the Hamiltonian matrix. The vibrational levels covering the range $\nu = 0$--$10$ were computed for each molecule at each magnetic field strength (see below), both with and without the inclusion of the DBOC contribution to the potential energy curve. Comprehensive computational results for each molecule are detailed in the subsequent dedicated sections.

\section{H$_2$ Molecule}
\label{h2}
\begin{figure}[h!]
\includegraphics[scale=0.95]{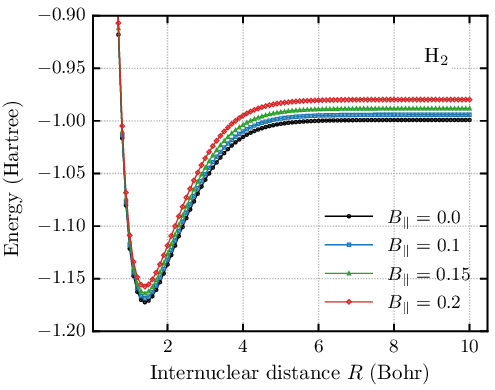}
\caption{Potential energy curves of H$_2$ molecule at different field strengths $B_{\parallel}$ (in atomic units), oriented parallel to the molecular axis.}
\label{fig:H2_PEC}
\end{figure}

The H$_2$ molecule is considered first, as it represents the simplest diatomic system and serves as a natural starting point for benchmarking the present approach. The corresponding potential energy curves (PECs) were calculated as a function of the magnetic field strength: the evaluation was performed at $B_{\parallel} = 0$~a.u., followed by calculations at $B_{\parallel} = 0.10$, $0.15$, and $0.20$~a.u., which correspond to strong magnetic field regimes. The resulting PECs are shown in Fig.~\ref{fig:H2_PEC} and are in good agreement with the data reported in Ref.~\cite{10.1063/5.0269984}.
Since the detailed calculation of these PECs has been recently described in Ref.~\cite{10.1063/5.0269984}, we omit the accompanying analysis here for the sake of brevity.

\begin{figure}[h!]
\includegraphics[scale=0.95]{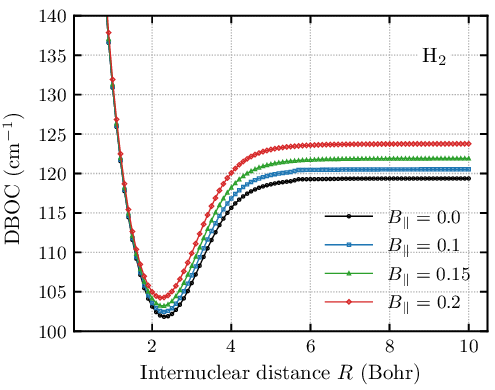}
\caption{Diagonal Born--Oppenheimer correction for the H$_2$ molecule at different magnetic field strengths $B_{\parallel}$ (in atomic units), oriented parallel to the molecular axis.}
\label{fig:H2_DBOC}
\end{figure}

Figure~\ref{fig:H2_DBOC} shows the DBOC for the H$_2$ molecule as a function of the internuclear distance (in bohr), with the correction given in cm$^{-1}$.
As follows from Fig.~\ref{fig:H2_DBOC}, in the repulsive region (at small separations), the correction exhibits a very weak dependence on the magnetic field.
As the internuclear distance approaches the equilibrium value, the DBOC curves for different magnetic field strengths follow a similar trend. However, at the equilibrium separation itself, their values begin to visibly diverge. Finally, at large distances, the asymptotic limit is reached at approximately $R=5$~bohr, which clearly illustrates the differences arising from the field variation.

Potential energies computed with and without the DBOC contribution for the H$_2$ molecule are compared in Table~\ref{tab:1a}.
\begin{table}[ht!]
\centering
\small
\caption{Potential energy curves of the H$_2$ molecule as a function of internuclear
distance $R$ (in bohr), computed with and without the DBOC at several magnetic field
strengths $B_{\parallel}$. Energies are in atomic units, the DBOC is given in
cm$^{-1}$.}
\begin{tabular}{ c c c c }
\hline
$R$ & Energy & Energy + DBOC & DBOC (cm$^{-1}$) \\
\hline
\multicolumn{4}{c}{$B_{\parallel}= 0.00$} \\
\hline
1.0 & -1.122\;127\;501 & -1.121\;531\;187 & 130.9 \\
2.0 & -1.136\;937\;088 & -1.136\;467\;162 & 103.1 \\
3.0 & -1.056\;400\;100 & -1.055\;916\;607 & 106.1 \\
4.0 & -1.015\;779\;968 & -1.015\;253\;031 & 115.6 \\
5.0 & -1.003\;329\;516 & -1.002\;788\;984 & 118.6 \\
\hline
\multicolumn{4}{c}{$B_{\parallel}= 0.10$} \\
\hline
1.0 & -1.118\;905\;146 & -1.118\;308\;191 & 131.0 \\
2.0 & -1.132\;381\;895 & -1.131\;909\;892 & 103.6 \\
3.0 & -1.051\;161\;561 & -1.050\;673\;560 & 107.1 \\
4.0 & -1.010\;555\;808 & -1.010\;023\;476 & 116.8 \\
5.0 & -0.998\;261\;865 & -0.997\;715\;975 & 119.8 \\
\hline
\multicolumn{4}{c}{$B_{\parallel}= 0.15$} \\
\hline
1.0 & -1.114\;915\;467 & -1.114\;317\;173 & 131.3 \\
2.0 & -1.126\;779\;905 & -1.126\;305\;268 & 104.2 \\
3.0 & -1.044\;754\;305 & -1.044\;260\;891 & 108.3 \\
4.0 & -1.004\;175\;673 & -1.003\;636\;970 & 118.2 \\
5.0 & -0.992\;064\;937 & -0.991\;512\;701 & 121.2 \\
\hline
\multicolumn{4}{c}{$B_{\parallel}= 0.20$} \\
\hline
1.0 & -1.109\;396\;894 & -1.108\;795\;827 & 131.9 \\
2.0 & -1.119\;091\;461 & -1.118\;613\;094 & 105.0 \\
3.0 & -1.036\;013\;498 & -1.035\;512\;828 & 109.9 \\
4.0 & -0.995\;485\;954 & -0.994\;938\;866 & 120.1 \\
5.0 & -0.983\;613\;438 & -0.983\;052\;832 & 123.0 \\
\hline
\end{tabular}
\label{tab:1a}
\end{table}

The correction is of the order of $10^{-4}$~a.u. Its magnitude is determined not by the nuclear masses alone, but rather by the product of the inverse nuclear mass and the sensitivity of the electronic wavefunction to nuclear displacement. For H$_2$ the second factor is small, since only two electrons follow the moving nuclei, whereas the first is the largest among the systems considered here. The correction is largest at short internuclear distances, passes through a shallow minimum near $R\approx 2.5$~bohr (well outside the equilibrium region), and then rises toward the dissociation limit, where it saturates at a value that increases by less than $4\%$ over the entire range of field strengths studied. This magnitude is comparable to the relativistic effects and energy differences now routinely resolved in high-precision spectroscopy of light molecules, meaning that omitting the DBOC would strictly limit the attainable accuracy for such systems.

\section{HeH$^{+}$ Molecule}
\label{HeH+}
\begin{figure}[h!]
\includegraphics[scale=1]{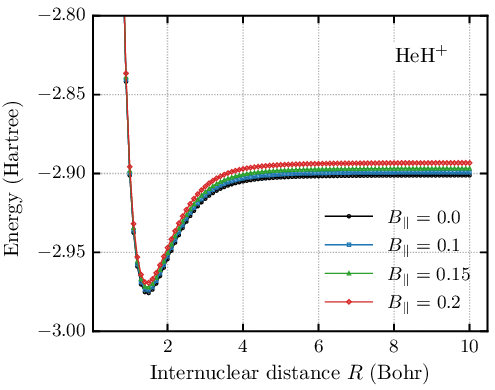}
\caption{Potential energy curves of HeH$^{+}$  molecule at different field strengths $B_{\parallel}$(in atomic units), oriented parallel to the molecular axis.}
\label{fig:HeH_PEC}
\end{figure}

In this section, the HeH$^{+}$ molecular ion is considered. This species is of profound interest in astrophysics, as it participated in the earliest chemical processes after the Universe became transparent~\cite{A.Dalgarno_2005} and is widely considered to be the first molecule ever formed. Moreover, the recent observation of the HeH$^{+}$ molecule in the NGC 7027 planetary nebula~\cite{NGC7027} has significantly revived research in this area. For this system, the computational procedure was identical to that described in Sec.~\ref{h2}. The resulting PECs are presented in Fig.~\ref{fig:HeH_PEC}.

\begin{figure}[h!]
\includegraphics[scale=1]{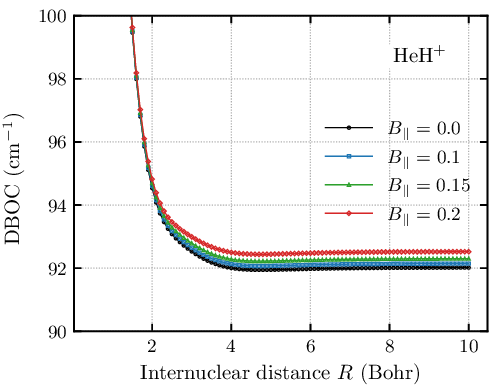}
\caption{DBOC for the HeH$^{+}$ molecule at different field strengths $B_{\parallel}$(in atomic units), oriented parallel to the molecular axis.}
\label{fig:HeH_DBOC}
\end{figure}

Figure~\ref{fig:HeH_DBOC} presents the corresponding results of DBOC calculations for the HeH$^{+}$ molecule.
In particular, the correction exhibits a behavior significantly different from that of H$_2$ discussed in the previous section. As before, the DBOC remains nearly insensitive to the magnetic field strength at short internuclear distances, where the curves calculated for different fields are virtually indistinguishable in the repulsive region. The most significant difference from H$_2$ is that the DBOC for HeH$^{+}$ does not exhibit a pronounced minimum; instead, the correction saturates almost monotonically with increasing internuclear distance, reaching its asymptotic limit at approximately $R=5$~bohr. This minimum, absent in both heteronuclear species studied here, is plausibly related to the symmetric charge distribution of a homonuclear system, although the present set of three molecules is too small to establish this as a general rule.

Table~\ref{tab:HeHp_PEC_DBOC} lists the potential energy curve energies alongside those augmented by the DBOC for the HeH$^{+}$ molecule at several magnetic field strengths $B_{\parallel}$. The correction is largest at the shortest internuclear distance considered ($114.3$~cm$^{-1}$ at $R=1$~bohr) and decreases monotonically with increasing $R$, leveling off at approximately $92$~cm$^{-1}$ for $R\gtrsim3$~bohr as the molecule approaches the dissociation limit.
At this limit, the electronic density of each fragment follows its own nucleus, and the DBOC saturates at its atomic value. This asymptotic value is roughly $20\%$ below that found for H$_2$, which at first sight may appear surprising given that the electrons of HeH$^{+}$ are bound substantially more tightly. The reason is that in H$_2$ both nuclei carry electrons that follow their motion, whereas in HeH$^{+}$ the proton is bare and only the helium fragment contributes. Per contributing atom the correction is indeed larger for HeH$^{+}$ ($87.4$~cm$^{-1}$ from the helium kinetic energy alone) than for hydrogen ($59.7$~cm$^{-1}$), but H$_2$ receives this contribution twice. The field dependence for HeH$^{+}$ is the weakest among the three systems studied in this work. As $B_{\parallel}$ is increased from $0.00$ to $0.20$~a.u., the plateau rises only from $91.9$ to $92.4$~cm$^{-1}$ (an increase of just $0.5\%$), compared with $4\%$ for H$_2$ and $2.5\%$ for LiH (see below).
\begin{table}[tb]
\centering
\small
\caption{Potential energy curves of the HeH$^{+}$ molecule as a function of internuclear
distance $R$ (in bohr), computed with and without the DBOC at several magnetic field
strengths $B_{\parallel}$. Energies are in atomic units. The DBOC itself is given in
cm$^{-1}$ in the last column. $R=1.5$~bohr is close to the equilibrium distance.}
\begin{tabular}{ c c c c }
\hline
$R$ & Energy & Energy + DBOC & DBOC (cm$^{-1}$) \\
\hline
\multicolumn{4}{c}{$B_{\parallel} = 0.00$} \\
\hline
1.0 & -2.901\;217\;963 & -2.900\;697\;119 & 114.3 \\
1.5 & -2.975\;730\;074 & -2.975\;276\;865 & 99.5 \\
2.0 & -2.954\;269\;955 & -2.953\;839\;215 & 94.5 \\
3.0 & -2.915\;976\;983 & -2.915\;555\;388 & 92.5 \\
4.0 & -2.904\;999\;926 & -2.904\;580\;721 & 92.0 \\
5.0 & -2.902\;456\;755 & -2.902\;037\;814 & 91.9 \\
\hline
\multicolumn{4}{c}{$B_{\parallel} = 0.10$} \\
\hline
1.0 & -2.899\;848\;034 & -2.899\;327\;115 & 114.3 \\
1.5 & -2.974\;101\;656 & -2.973\;648\;264 & 99.5 \\
2.0 & -2.952\;467\;809 & -2.952\;036\;746 & 94.6 \\
3.0 & -2.914\;024\;334 & -2.913\;602\;209 & 92.6 \\
4.0 & -2.903\;017\;559 & -2.902\;597\;784 & 92.1 \\
5.0 & -2.900\;469\;363 & -2.900\;049\;849 & 92.1 \\
\hline
\multicolumn{4}{c}{$B_{\parallel} = 0.15$} \\
\hline
1.0 & -2.898\;138\;453 & -2.897\;617\;440 & 114.3 \\
1.5 & -2.972\;070\;920 & -2.971\;617\;300 & 99.6 \\
2.0 & -2.950\;221\;952 & -2.949\;790\;484 & 94.7 \\
3.0 & -2.911\;593\;128 & -2.911\;170\;342 & 92.8 \\
4.0 & -2.900\;549\;973 & -2.900\;129\;493 & 92.3 \\
5.0 & -2.897\;995\;652 & -2.897\;575\;430 & 92.2 \\
\hline
\multicolumn{4}{c}{$B_{\parallel} = 0.20$} \\
\hline
1.0 & -2.895\;750\;263 & -2.895\;229\;115 & 114.4 \\
1.5 & -2.969\;236\;677 & -2.968\;782\;734 & 99.6 \\
2.0 & -2.947\;090\;225 & -2.946\;658\;188 & 94.8 \\
3.0 & -2.908\;206\;873 & -2.907\;783\;168 & 93.0 \\
4.0 & -2.897\;114\;130 & -2.896\;692\;673 & 92.5 \\
5.0 & -2.894\;551\;473 & -2.894\;130\;274 & 92.4 \\
\hline
\end{tabular}
\label{tab:HeHp_PEC_DBOC}
\end{table}

\section{LiH Molecule}
\label{LiH}
The LiH molecule also plays a crucial role in the chemistry of the early Universe~\cite{A.Dalgarno_2005}. As has been shown, this species contributed significantly to the cooling of primordial gas and the subsequent formation of the first cosmic structures~\cite{Bovino_2011,LiH_EarlyUniverse}. However, detecting LiH near compact astrophysical objects remains a major observational challenge due to its relatively low cosmic abundance. Despite this difficulty, the potential existence of LiH in the atmospheres of highly magnetized white dwarfs has attracted considerable interest. Because the molecule in such extreme environments is exposed to strong magnetic fields, evaluating its properties under these conditions directly motivates the present study.

\begin{figure}[h!]
\includegraphics[scale=1]{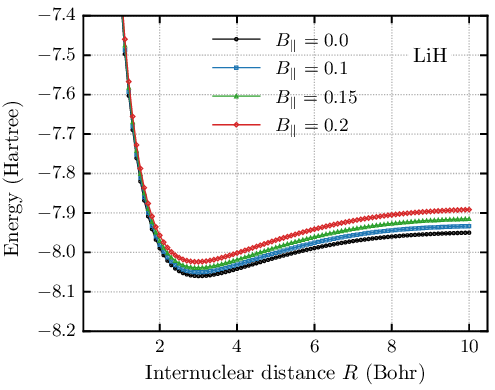}
\caption{Potential energy curves of the LiH molecule at different field strengths $B_{\parallel}$ (in atomic units), oriented parallel to the molecular axis.}
\label{fig:LiH_PEC}
\end{figure}

Figure~\ref{fig:LiH_PEC} presents the calculated potential energy curves for the LiH system at magnetic field strengths $B_{\parallel} = 0.0$, $0.10$, $0.15$, and $0.20$~a.u. The ground-state PECs exhibit a clear potential well with a pronounced minimum, confirming the stability of the bound molecular state. Both the depth and the equilibrium position of this minimum shift systematically with increasing magnetic field strength, in full agreement with the findings of Ref.~\cite{10.1063/5.0269984}.

\begin{figure}[h!]
\includegraphics[scale=1]{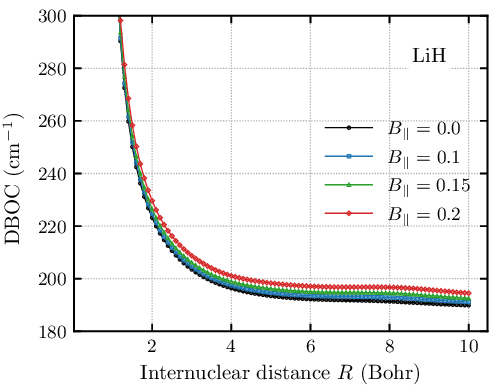}
\caption{Diagonal Born--Oppenheimer correction for the LiH molecule at different magnetic field strengths $B_{\parallel}$ (in atomic units), oriented parallel to the molecular axis.}
\label{fig:LiH_DBOC}
\end{figure}

The corresponding DBOC results for LiH are shown in Fig.~\ref{fig:LiH_DBOC}.
In contrast to H$_2$ and similarly to HeH$^{+}$, the correction does not feature a minimum before leveling off. Instead, it decreases monotonically with increasing internuclear distance, tending to a constant asymptotic value at large separations. This behavior can be attributed to the strongly ionic character of the LiH bond and the resulting asymmetric shape of its potential energy curve, which leads to a more gradual variation of the electronic wavefunction with nuclear displacement.

Numerical results for the PECs of LiH computed with and without the DBOC contribution are
presented in Table~\ref{tab:3a}.
\begin{table}[h!]
\centering
\small
\caption{Potential energy curves of the LiH molecule as a function of internuclear
distance $R$ (in bohr), computed with and without the DBOC at several magnetic field
strengths $B_{\parallel}$. Energies are in atomic units. The DBOC itself is given in
cm$^{-1}$ in the last column. $R=3.0$~bohr is the equilibrium distance.}
\begin{tabular}{ c c c c }
\hline
$R$ & Energy & Energy + DBOC & DBOC (cm$^{-1}$) \\
\hline
\multicolumn{4}{c}{$B_{\parallel}=0.00$} \\
\hline
1.0 & -7.367\;847\;575 & -7.366\;134\;443 & 376.0 \\
2.0 & -7.989\;396\;352 & -7.988\;379\;468 & 223.2 \\
3.0 & -8.060\;020\;395 & -8.059\;093\;276 & 203.5 \\
4.0 & -8.041\;876\;958 & -8.040\;982\;265 & 196.4 \\
5.0 & -8.013\;392\;854 & -8.012\;511\;586 & 193.4 \\
10.0 & -7.950\;104\;908 & -7.949\;239\;853 & 189.9 \\
\hline
\multicolumn{4}{c}{$B_{\parallel}=0.10$} \\
\hline
1.0 & -7.356\;689\;714 & -7.354\;994\;338 & 372.1 \\
2.0 & -7.980\;797\;680 & -7.979\;774\;610 & 224.5 \\
3.0 & -8.050\;403\;108 & -8.049\;470\;930 & 204.6 \\
4.0 & -8.030\;996\;977 & -8.030\;097\;180 & 197.5 \\
5.0 & -8.001\;234\;739 & -8.000\;347\;768 & 194.7 \\
10.0 & -7.933\;377\;189 & -7.932\;506\;299 & 191.1 \\
\hline
\multicolumn{4}{c}{$B_{\parallel}=0.15$} \\
\hline
1.0 & -7.343\;864\;137 & -7.342\;194\;288 & 366.5 \\
2.0 & -7.970\;565\;974 & -7.969\;534\;230 & 226.4 \\
3.0 & -8.039\;024\;554 & -8.038\;085\;294 & 206.1 \\
4.0 & -8.018\;299\;937 & -8.017\;393\;491 & 198.9 \\
5.0 & -7.987\;288\;500 & -7.986\;394\;501 & 196.2 \\
10.0 & -7.914\;882\;559 & -7.914\;005\;080 & 192.6 \\
\hline
\multicolumn{4}{c}{$B_{\parallel}=0.20$} \\
\hline
1.0 & -7.327\;422\;979 & -7.325\;778\;382 & 360.9 \\
2.0 & -7.956\;966\;665 & -7.955\;920\;454 & 229.6 \\
3.0 & -8.023\;968\;915 & -8.023\;017\;831 & 208.7 \\
4.0 & -8.001\;689\;732 & -8.000\;773\;463 & 201.1 \\
5.0 & -7.969\;305\;701 & -7.968\;402\;133 & 198.3 \\
10.0 & -7.891\;615\;494 & -7.890\;729\;402 & 194.5 \\
\hline
\end{tabular}
\label{tab:3a}
\end{table}
The correction is largest at short internuclear distances, reaching $376$~cm$^{-1}$ ($1.7\times10^{-3}$~a.u.) at $R=1$~bohr, and decreases with increasing $R$ to about $190$~cm$^{-1}$ ($0.9\times10^{-3}$~a.u.) at $R=10$~bohr. This is roughly twice the value found for HeH$^{+}$, reflecting the larger number of electrons that follow the moving nuclei rather than any difference in nuclear masses, even though the Li nucleus is the heaviest considered here.
In contrast to H$_2$, the field dependence changes sign along the curve: at $R=1$~bohr, the correction decreases by about $4\%$ between $B_{\parallel}=0.00$ and $0.20$~a.u., whereas for $R\gtrsim 2$~bohr, it increases by approximately $2.5\%$ over the same range. The sign change occurs near $R\approx 1.5$~bohr, as seen in Fig.~\ref{fig:LiH_DBOC}, where the curves for different field strengths intersect at this internuclear distance.

\section{Vibrational Structure Calculations}
\label{vibrational}

In this section, we present the results for the vibrational spectra of the H$_2$, HeH$^{+}$, and LiH molecules subjected to strong magnetic fields. The vibrational energy levels were determined using the \texttt{Duo} software by solving the one-dimensional nuclear Schr\"odinger equation. For each molecule, we utilized the potential energy curves obtained in the previous section across various field strengths. To clearly isolate the effect of the diagonal Born-Oppenheimer correction, the computations were performed twice: both with and without the inclusion of the DBOC. Specifically, the first eleven vibrational energy levels ($\nu = 0$--$10$) were calculated for each system, with all values reported relative to the ground state ($\nu=0$). The resulting spectroscopic data for the hydrogen molecule are compiled in Table~\ref{tab:4a}.
\begin{table}[h!]
\centering
\caption{The first eleven vibrational energy levels of  H$_2$ molecule (in cm$^{-1}$) for various field strengths $B_{\parallel}$ (in atomic units). The second row in each cell corresponds to the value of the vibrational level taking into account the DBOC.}
\begin{tabular}{ c  c  c  c  c }
\hline
\multicolumn{5}{c}{H$_2$}  \\
\hline
$B_{\parallel}$ & 0.0 & 0.1 & 0.15 & 0.2\\
\hline
$\nu$ = 0 & 0 & 0 & 0 & 0 \\
$\nu$ = 1 & 4156.59	& 4182.36	& 4213.22	& 4254.30\\
  & 4155.23	& 4181.04	& 4211.95	& 4253.08\\
$\nu$ = 2 & 8077.18	& 8128.23	& 8189.32	& 8270.56\\
  & 8074.68	& 8125.82	& 8187.00	& 8268.35\\
$\nu$ = 3 & 11766.71	& 11842.45	& 11933.01	& 12053.34\\
  & 11763.34	& 11839.20	& 11929.90	& 12050.40\\
$\nu$ = 4 & 15228.01	& 15327.71	& 15446.83	& 15605.00\\
  & 15224.02	& 15323.89	& 15443.20	& 15601.61\\
$\nu$ = 5 & 18461.64	& 18584.43	& 18731.04	& 18925.58\\
  & 18457.31	& 18580.30	& 18727.15	& 18921.99\\
$\nu$ = 6 & 21465.82	& 21610.63	& 21783.44	& 22012.63\\
  & 21461.41	& 21606.48	& 21779.59	& 22009.13\\
$\nu$ = 7 & 24236.15	& 24401.71	& 24599.19	& 24860.96\\
  & 24231.96	& 24397.84	& 24595.66	& 24857.86\\
$\nu$ = 8 & 26765.13	& 26949.93	& 27170.25	& 27462.19\\
  & 26761.46	& 26946.62	& 27167.34	& 27459.78\\
$\nu$ = 9 & 29041.33	& 29243.53	& 29484.53	& 29803.80\\
  & 29038.46	& 29241.08	& 29482.55	& 29802.38\\
$\nu$ = 10 & 31048.13	& 31265.57	& 31524.69	& 31867.91\\
  & 31046.34	& 31264.26	& 31523.90	& 31867.77\\
\hline
\end{tabular}
\label{tab:4a}
\end{table}

The zero-field vibrational levels of H$_2$ listed in Table~\ref{tab:4a} agree well with available theoretical~\cite{Vibrationo_H} and experimental data~\cite{article}. Furthermore, the vibrational energies obtained across various magnetic fields excluding the diagonal BO correction are in agreement with our previous results~\cite{10.1063/5.0269984}, with a relative deviation not exceeding $0.3\%$. To better assess the impact of the DBOC on the vibrational levels, one can examine the difference between the vibrational energies computed without (the values in the upper rows) and with the DBOC (the values in the lower rows) for each magnetic field strength. Accordingly, this DBOC-induced shift is evaluated as
\begin{eqnarray}
\label{vDBOC}
    \Delta E = E_{\nu}-E^{\mathrm{DBOC}}_{\nu}.
\end{eqnarray}
Here, $E_{\nu}$ denotes the vibrational energy (in cm$^{-1}$) of the level $\nu$ computed in the absence of the DBOC, while $E^{\mathrm{DBOC}}_{\nu}$ is the corresponding value with the correction included.

\begin{figure}[h!]
\includegraphics[scale=1]{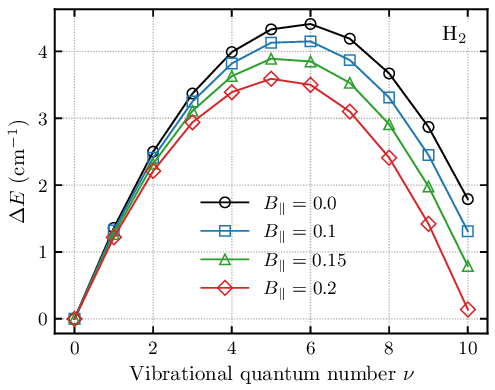}
\caption{Difference between vibrational energy levels computed without the DBOC and with the DBOC for the H$_2$ molecule at different magnetic field strengths $B_{\parallel}$ (in atomic units), oriented parallel to the molecular axis.}
\label{fig:H2_vibrational}
\end{figure}

The energy differences computed according to Eq.~\eqref{vDBOC} are presented graphically in Fig.~\ref{fig:H2_vibrational}.

In particular, Figure~\ref{fig:H2_vibrational} displays the $\Delta E$ values for the H$_2$ molecule, revealing a maximum shift at intermediate vibrational levels around $\nu \approx 5$--$6$ depending on the field strength. The appearance of an extremum in the diagonal Born--Oppenheimer correction as a function of the vibrational quantum number can be understood by comparing the potential energy curve with the DBOC profile plotted against the internuclear separation. According to Figs.~\ref{fig:H2_PEC} and~\ref{fig:H2_DBOC}, both of these curves exhibit a distinct minimum. Notably, the minimum of the DBOC is shifted relative to that of the PEC by approximately $1$~bohr across all investigated field strengths. The repulsive part of the DBOCs decreases almost linearly up to the PEC minima and remains essentially independent of the magnetic field strength. This manifests as a linear dependence of $\Delta E$ on the vibrational quantum number up to about $\nu=3$, see Fig.~\ref{fig:H2_vibrational}. Starting already within the DBOC minimum region in Fig.~\ref{fig:H2_DBOC}, which corresponds to $\nu\gtrsim 4$, a clear dependence on the field strength becomes apparent, with the initial decrease in the correction values being replaced by a subsequent increase. As indicated by Table~\ref{tab:4a}, the vibrational energy shifts vary around $4$~cm$^{-1}$ when passing through the DBOC minimum region. Thereafter, as the DBOC values grow, these shifts reach the maximum at $\nu \approx 5$--$6$ before gradually decreasing. Such a trend confirms that the DBOC minimum is displaced from the PEC well, meaning that the subsequent rapid growth of the correction takes place outside the quantization region. Consequently, the bound vibrational states exhibit diminished sensitivity to this growth as their wavefunctions decay toward the dissociation limit.

A notable conclusion to be drawn from Fig.~\ref{fig:H2_vibrational} is that for certain transitions, the impact of the forced nuclear motion on the vibrational energies can be minimized. While inherently small (on the order of a few cm$^{-1}$), the DBOC contribution diminishes significantly depending on the vibrational quantum number $\nu$, as exemplified by the $\nu=0$ and $\nu=10$ states at $B_{\parallel}=0.2$~a.u. Consequently, this contrast opens new avenues for probing strong magnetic fields, enabling a clear disentanglement of the static electronic potential energy curve deformation from the dynamical non-adiabatic effects.

On the one hand, transitions for which the DBOC nearly cancels, such as $\nu=0 \to \nu=10$ at $B_{\parallel}=0.2$~a.u., carry the spectral shift induced by the field-deformed electronic PEC almost in isolation. This cancellation arises only at this particular field strength and therefore does not by itself supply a field-independent reference. It does mean, however, that at this field the position of the $\nu=10$ level probes the electronic response alone. On the other hand, transitions close to the maximum of $\Delta E$, exemplified by $\nu=0 \to \nu=5$, are the most sensitive to the forced nuclear motion. Since the two contributions depend on $\nu$ in different ways, measuring several transitions within the same vibrational progression would in principle allow the static electronic deformation and the non-adiabatic contribution to be separated, without either having to be known in advance. This contrast provides a diagnostic handle on the interplay between electronic potential deformation and forced nuclear motion in strong, uniform magnetic fields.

Table~\ref{tab:4a} shows how strongly the magnetic field stretches the vibrational ladder of H$_2$. Between $B_{\parallel}=0$ and $0.2$~a.u., the fundamental vibrational interval increases from $4156.6$ to $4254.3$~cm$^{-1}$, while the $\nu=0 \to \nu=10$ total interval expands from $31048$ to $31868$~cm$^{-1}$, both representing an enhancement of approximately $2.5\%$. This structural stretching amounts to tens and hundreds of cm$^{-1}$. Crucially, the non-adiabatic DBOC contribution to these intervals never exceeds $4.4$~cm$^{-1}$ (for the $\nu=6$ level) at any field strength studied, indicating that the observed spectral changes are overwhelmingly electronic in origin.

Table~\ref{tab:4a} clearly reveals that this field-induced response is highly non-linear with respect to two distinct parameters: the external magnetic field strength and the vibrational quantum number $\nu$. The first, field-dependent non-linearity is demonstrated by comparing the field-induced energy shifts over two consecutive, equal intervals of field enhancement: from $B_{\parallel}=0.10$~a.u. to $0.15$~a.u., and subsequently from $0.15$~a.u. to $0.20$~a.u. Specifically, for the first interval, the shifts amount to $|\Delta E_{\nu=5}^{0.10 \to 0.15}| = 146.85$~cm$^{-1}$ and $|\Delta E_{\nu=10}^{0.10 \to 0.15}| = 259.64$~cm$^{-1}$, whereas for the second interval (the region of higher field strength), the corresponding displacements increase to $|\Delta E_{\nu=5}^{0.15 \to 0.20}| = 194.84$~cm$^{-1}$ and $|\Delta E_{\nu=10}^{0.15 \to 0.20}| = 343.87$~cm$^{-1}$. This confirms that the observed $24\%$ relative asymmetry in the level shifts stems directly from the non-linear scaling of the potential energy curve as a function of the magnetic field strength.

In turn, the physical origin of the second, state-to-state non-linearity lies in the state-specific nature of the molecular response. This non-linear scaling with respect to the vibrational quantum number $\nu$, as exemplified by the distinct absolute shifts for the $\nu=5$ and $\nu=10$ levels, is governed by the unique spatial localization of each vibrational wavefunction. Because different wavefunctions $\psi_{\nu}$ are confined to distinct regions of the field-deformed potential, the resulting energy levels undergo non-uniform displacements rather than a rigid, uniform shift. This non-linear behavior of the spectral shifts within a single vibrational progression directly reflects the non-linear, coordinate-dependent distortion of the electronic PEC induced by the strong, uniform magnetic field.

\begin{table}[h!]
\centering
\caption{The first eleven vibrational energy levels of HeH$^{+}$ molecule (in cm$^{-1}$) for various field strengths $B_{\parallel}$ (in atomic units). The second row in each cell corresponds to the value of the vibrational level taking into account the DBOC }
\begin{tabular}{ c  c  c  c  c }
\hline
\multicolumn{5}{c}{HeH$^{+}$ }  \\
\hline
$B_{\parallel}$ & 0.0 & 0.1 & 0.15 & 0.2\\
\hline
$\nu$ = 0 & 0 & 0	& 0	& 0\\
$\nu$ = 1 & 2902.87	& 2911.23	& 2921.56	& 2935.82\\
  & 2903.66	& 2912.04	& 2922.40	& 2936.69\\
$\nu$ = 2 & 5501.56	& 5518.29	& 5538.95	& 5567.45\\
  & 5503.41	& 5520.17	& 5540.88	& 5569.44\\
$\nu$ = 3 & 7793.60	& 7818.71	& 7849.72	& 7892.47\\
  & 7796.72	& 7821.89	& 7852.97	& 7895.82\\
$\nu$ = 4 & 9773.66	& 9807.19	& 9848.58	& 9905.60\\
  & 9778.29	& 9811.89	& 9853.38	& 9910.54\\
$\nu$ = 5 & 11433.47 & 11475.45	& 11527.27	& 11598.63\\
  & 11439.76	& 11481.85	& 11533.80	& 11605.34\\
$\nu$ = 6 & 12761.68	& 12812.17	& 12874.48	& 12960.25\\
  & 12769.72	& 12820.36	& 12882.83	& 12968.84\\
$\nu$ = 7 & 13745.96	& 13804.98	& 13877.80	& 13978.01\\
  & 13755.81	& 13815.00	& 13888.02	& 13988.51\\
$\nu$ = 8 & 14382.03	& 14449.23	& 14532.13	& 14646.25\\
  & 14393.67	& 14461.06	& 14544.20	& 14658.64\\
$\nu$ = 9 &  14702.83	& 14776.53	& 14867.51	& 14992.86\\
  & 14716.01	& 14789.93	& 14881.17	& 15006.88\\
$\nu$ = 10 & 14821.14	& 14898.01	& 14993.01	& 15124.07\\
  & 14835.07	& 14912.19	& 15007.50	& 15138.99\\
\hline
\end{tabular}
\label{tab:5a}
\end{table}

\begin{figure}[h!]
\includegraphics[scale=1]{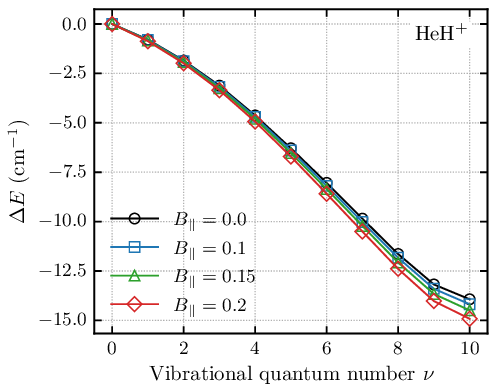}
\caption{Difference between vibrational energy levels computed without the DBOC and with the DBOC for the HeH$^{+}$ molecule at different magnetic field strengths $B_{\parallel}$ (in atomic units), oriented parallel to the molecular axis.}
\label{fig:HeH_vibrational}
\end{figure}

Table~\ref{tab:5a} compiles the results obtained for the vibrational energy levels of the HeH$^{+}$ molecule.
The numerical values for the lowest vibrational level at $B_{\parallel}=0.0$~a.u. agree well with the theoretical and experimental results reported in Refs.~\cite{10.1063/5.0269984,PhysRevLett.96.233002,10.1063/1.4768169_HeH+}. As shown in Fig.~\ref{fig:HeH_vibrational}, the energy shift $\Delta E$ exhibits an almost linear behavior, except at the bottom of the potential well (corresponding to $\nu=0,1$) and near the dissociation threshold ($\nu=9,10$).
This trend originates from the fact that the DBOC curve for HeH$^{+}$, see Fig.~\ref{fig:HeH_DBOC}, lacks a potential well. Instead, it is a strictly monotonic function that approaches its asymptotic limit at large internuclear distances, unlike in H$_2$. The fact that the DBOC decreases almost linearly within the region of the adiabatic potential energy well directly translates into a characteristic linear dependence on the vibrational quantum number, which is clearly illustrated in Fig.~\ref{fig:HeH_vibrational}. Another evident conclusion from Fig.~\ref{fig:HeH_vibrational} is that the DBOC exhibits a very weak dependence on the magnetic field strength, mirroring the behavior shown in Fig.~\ref{fig:HeH_DBOC}. A noticeable deviation occurs only for high $\nu$ states, directly reflecting the field-dependent divergence of the DBOC curves at large internuclear distances.

For HeH$^{+}$ the field-induced stretching of the vibrational ladder is somewhat weaker than for H$_2$: between $B_{\parallel}=0$ and $0.2$~a.u. the fundamental interval grows from $2902.9$ to $2935.8$~cm$^{-1}$ ($1.1\%$) and the $\nu=0\to\nu=10$ interval from $14821$ to $15124$~cm$^{-1}$ ($2.0\%$). Nevertheless, both the field-dependent non-linearity and the non-linear scaling with respect to the vibrational quantum number $\nu$ remain clearly pronounced in this system. The DBOC contribution increases almost linearly with $\nu$, from $0.8$~cm$^{-1}$ at $\nu=1$ to about $14$~cm$^{-1}$ at $\nu=10$, and depends only weakly on the field, mirroring the behavior of Fig.~\ref{fig:HeH_DBOC}. Relative to the level spacings it is thus considerably larger than in H$_2$, and for the highest levels it is no longer negligible at the accuracy currently attainable for molecular line positions in stellar plasmas.

\begin{table}[h!]
\centering
\caption{The first eleven vibrational energy levels of the LiH molecule (in cm$^{-1}$) for various field strengths $B_{\parallel}$ (in atomic units). The second row in each cell corresponds to the value of the vibrational level taking into account the DBOC}
\begin{tabular}{ c  c  c  c  c }
\hline
\multicolumn{5}{c}{LiH}  \\
\hline
$B_{\parallel}$ & 0.0 & 0.1 & 0.15 & 0.2\\
\hline
$\nu$ = 0 & 0 & 0 & 0 & 0 \\
$\nu$ = 1 & 1359.77	& 1386.10	& 1412.91	& 1443.85\\
  & 1359.02	& 1385.37	& 1412.16	& 1443.03\\
$\nu$ = 2 & 2675.73	& 2728.79	& 2782.53	& 2844.28\\
  & 2674.28	& 2727.36	& 2781.08	& 2842.69\\
$\nu$ = 3 &  3949.26	& 4029.50	& 4110.37	& 4202.83\\
  & 3947.14	& 4027.41	& 4108.26	& 4200.54\\
$\nu$ = 4 & 5181.57	& 5289.49	& 5397.75	& 5520.93\\
  & 5178.79	& 5286.77	& 5395.01	& 5517.98\\
$\nu$ = 5 & 6373.67	& 6509.85	& 6645.82	& 6799.82\\
  & 6370.26	& 6506.52	& 6642.48	& 6796.23\\
$\nu$ = 6 & 7526.47	& 7691.53	& 7855.60	& 8040.57\\
  & 7522.46	& 7687.63	& 7851.69	& 8036.40\\
$\nu$ = 7 & 8640.80	& 8835.44	& 9028.02	& 9244.19\\
  & 8636.21	& 8830.97	& 9023.57	& 9239.45\\
$\nu$ = 8 &  9717.42	& 9942.38	& 10163.97	& 10411.59\\
  & 9712.28	& 9937.39	& 10159.00	& 10406.32\\
$\nu$ = 9 & 10757.04	& 11013.13	& 11264.24	& 11543.65\\
  & 10751.36	& 11007.62	& 11258.77	& 11537.86\\
$\nu$ = 10 & 11760.30	& 12048.39	& 12329.62	& 12641.19\\
  & 11754.10	& 12042.40	& 12323.68	& 12634.91\\
\hline
\end{tabular}
\label{tab:6a}
\end{table}

\begin{figure}[h!]
\includegraphics[scale=1]{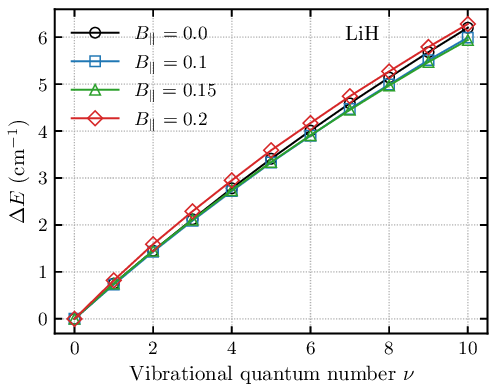}
\caption{Difference between vibrational energy levels computed without the DBOC and with the DBOC for the LiH molecule at different magnetic field strengths $B_{\parallel}$ (in atomic units), oriented parallel to the molecular axis.}
\label{fig:LiH_vibrational}
\end{figure}

The vibrational energy levels of LiH are listed in Table~\ref{tab:6a}. The values
obtained without the DBOC are consistent with those reported in
Ref.~\cite{10.1063/5.0269984}. The DBOC shift $\Delta E_\nu$ grows almost linearly with
the vibrational quantum number, from $0.75$ to $6.20$~cm$^{-1}$ between $\nu=1$ and
$\nu=10$ at zero field, a straight line reproducing the entire progression to within
$0.3$~cm$^{-1}$ at all field strengths studied. Similar to the HeH$^{+}$ case, the DBOC contribution to these vibrational levels exhibits a very weak dependence on the external magnetic field strength. This trend originates directly from the shape of the DBOC curve plotted against the internuclear separation, see Fig.~\ref{fig:LiH_DBOC}. For clarity, the corresponding energy shifts across different field strengths are illustrated in Fig.~\ref{fig:LiH_vibrational}.

First, the DBOC profile of LiH, Fig.~\ref{fig:LiH_DBOC}, continues to decrease
gradually throughout the investigated range of internuclear distances and has not yet
saturated at $R=10$~bohr, the decrease persisting well beyond the region of the PEC
minimum. Second, the adiabatic potential well of LiH is considerably broader and
shallower than those of the previously discussed molecules, Fig.~\ref{fig:LiH_PEC},
and its minimum falls at an internuclear separation comparable to the inflection point
of the DBOC curve. Third, LiH is by far the least anharmonic of the three systems: the
level spacings contract from $1359.8$ to $1003.3$~cm$^{-1}$ between $\nu=0\to1$ and
$\nu=9\to10$ at zero field, a reduction of only $26\%$, whereas the corresponding
figures are $52\%$ for H$_2$ and $96\%$ for HeH$^{+}$. The vibrational ladder of LiH
therefore remains close to equidistant over the whole range considered.

The interplay of these three factors (namely, the monotonic decrease of the DBOC, the close alignment between the PEC minimum and the DBOC inflection point, and the near-harmonic character of the vibrational ladder) ultimately dictates the linear dependence of the diagonal Born--Oppenheimer correction to the vibrational energies on $\nu$. Furthermore, while an increase in $\nu$ for the HeH$^{+}$ molecule drives the energy shifts toward larger negative values, the opposite trend is observed here, manifesting as a growth in positive values and thus demonstrating an inverted behavior relative to the HeH$^{+}$ species.

As in the previous cases, the weak field dependence within the repulsive region of the DBOC is clearly reflected in the behavior of the $\nu=0,1,2$ states in Fig.~\ref{fig:LiH_vibrational}. Minor discrepancies emerge starting from $\nu \approx 2$, in strict accordance with the spatial divergence of the DBOC curves plotted against the internuclear separation in Fig.~\ref{fig:LiH_DBOC}. 
Finally, LiH exhibits the largest field-induced stretching of the three systems: between $B_{\parallel}=0$ and $0.2$~a.u. the fundamental interval grows from $1359.8$ to $1443.9$~cm$^{-1}$ ($6.2\%$) and the $\nu=0\to\nu=10$ interval from $11760$ to $12641$~cm$^{-1}$ ($7.5\%$). This reflects the broad and shallow potential well of LiH, Fig.~\ref{fig:LiH_PEC}, whose depth and equilibrium position are both displaced appreciably already at moderate fields. The DBOC shift grows nearly linearly with $\nu$ and reaches $6.3$~cm$^{-1}$ at $\nu=10$, again with only a weak dependence on the field strength.

While modern laboratory experiments measure the H$_2$ vibrational quanta with an exceptional precision of $10^{-3}$--$10^{-7}$~cm$^{-1}$~\cite{Campargue_H2_2011,HT_Precision_2024}, observational limitations in stellar plasma environments, such as the atmospheres of magnetic white dwarfs, restrict the frequency determination accuracy to approximately $0.1$--$2.0$~cm$^{-1}$~\cite{Parigger_WhiteDwarf_Plasma,JWST_H2_MIRI_2026}. The DBOC-induced shifts obtained here range from a few tenths to about $15$~cm$^{-1}$, and the field-induced changes of the vibrational intervals themselves amount to hundreds of cm$^{-1}$, so that both exceed these empirical margins. Whether these shifts can be resolved in practice will nevertheless depend on the line broadening in the source, since molecular lines in the atmospheres of magnetic white dwarfs are broadened by the inhomogeneity of the field across the stellar disk and by pressure, and the molecular orientation is not fixed, whereas the present calculation is restricted to $J=0$ and to a strictly parallel geometry. Since the DBOC itself varies only weakly with the field, its inclusion is required for the absolute positions of the lines rather than for their field-induced displacements. In this context, it is worth noting the growing role of the DBOC for transitions originating from highly excited vibrational states in heteronuclear species. While for the hydrogen molecule the DBOC remains on the order of $1$~cm$^{-1}$, which is close to the observational resolution limit, it nevertheless reveals a fundamentally distinct behavior inherent to homonuclear diatomic systems.

\section{Conclusion}
\label{Conclusion}

We have investigated the diagonal Born--Oppenheimer correction (DBOC) for the light diatomic systems H$_2$, HeH$^+$, and LiH at the Hartree--Fock and CISD levels of theory in parallel magnetic fields $B_\parallel = 0.0$--$0.2$~a.u. For H$_2$ and HeH$^+$, the DBOC is on the order of $10^{-4}$~a.u., whereas for LiH it is slightly larger, $\sim 10^{-3}$~a.u., primarily reflecting the increased number of electrons accompanying nuclear motion rather than differences in nuclear masses.

For H$_2$, the DBOC attains its maximum at short internuclear distances, decreases to a shallow minimum near $R \approx 2.5$~bohr (well beyond the equilibrium region), and subsequently increases toward the dissociation limit, where it approaches a nearly constant asymptotic value. In contrast, for HeH$^{+}$, the correction decreases monotonically with increasing $R$ and gradually saturates at large separations. A similar monotonic behavior is observed for LiH, where the DBOC decreases smoothly without exhibiting a minimum. However, due to the strongly ionic character of the LiH bond, its asymptotic limit is not reached even at $R=10$~bohr.

\begin{figure}[h!]
\includegraphics[scale=0.9]{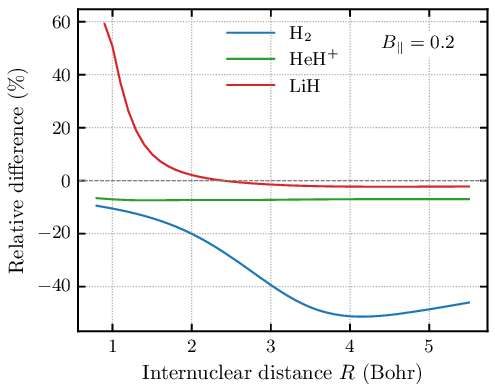}
\caption{
Relative difference between the DBOCs (in \%) computed at the UHF and UCISD levels of theory as a function of the internuclear distance for H$_2$, HeH$^{+}$, and LiH in a parallel magnetic field $B_{\parallel}=0.2$~a.u. 
}
\label{fig:DBOC_reldiff}
\end{figure}

A separate point of discussion is the effect of electron correlation. Comparing the unrestricted Hartree--Fock (UHF) and unrestricted CISD (UCISD) levels indicates that its impact on the DBOC is molecule-specific and most significant at short internuclear distances under a magnetic field strength of $0.2$~a.u. Figure~\ref{fig:DBOC_reldiff} illustrates the relative difference between the DBOC values computed at the UHF and UCISD levels, $(\mathrm{DBOC}_{\mathrm{UHF}}-\mathrm{DBOC}_{\mathrm{UCISD}})/\mathrm{DBOC}_{\mathrm{UHF}}$, as a function of the internuclear separation.

The deviations observed for these three systems stem from distinct physical origins and manifest in different coordinate regions. For LiH, the pronounced peak at $R<3$~bohr reflects the enhanced sensitivity of the DBOC to electron correlation in this highly polar system. We relate this to the strongly ionic character of the LiH bond: when the nuclei are pushed closer than the equilibrium distance, the electron density shifts rapidly between covalent- and ionic-like configurations. A single UHF determinant describes this shift poorly, whereas UCISD, which includes electron correlation, effectively captures this transition. For H$_2$, the largest deviation occurs at large internuclear separations, reaching approximately $-45\%$ near $R = 4$~bohr. In this region, a single-determinant reference fails to describe bond dissociation, and the UHF wavefunction ceases to be a spin eigenstate. In contrast, UCISD, which is equivalent to full configuration interaction for a two-electron system, remains strictly accurate throughout the entire range of distances. Finally, HeH$^{+}$ dissociates into a closed-shell He atom and a bare proton, thereby exhibiting no such breakdown and consistently remaining near $-7\%$ over the whole range. This suggests a general rule: a correlated wavefunction is required for highly polar systems at short range, as well as for any system whose unrestricted reference breaks spin symmetry upon dissociation, while UHF may still be adequate for weakly polar diatomics near equilibrium. This distinction becomes even more important in the presence of an external magnetic field, since the field can further increase the sensitivity of the electronic response to the level of theory employed.


A second key result concerns how the DBOC responds to the magnetic field itself. The DBOC-induced shift of the vibrational levels, $\Delta E = E_{\nu} - E^{\mathrm{DBOC}}_{\nu}$, Eq.~\eqref{vDBOC}, ranges from a few tenths to approximately $15$~cm$^{-1}$ in magnitude, growing with $B_\parallel$ for HeH$^{+}$ and LiH while decreasing for H$_2$. For HeH$^{+}$ and LiH, this growth mainly follows the general stretching of the vibrational level spacings caused by the field, rather than a direct coupling between the DBOC and $B_\parallel$. Indeed, when $\Delta E$ is expressed relative to the bare vibrational energy, it remains close to its zero-field value at every field strength studied, changing by no more than a few percent. For H$_2$, this relative weight decreases somewhat more, by $10$--$20\%$ for the lower and intermediate levels. In other words, over the range of $B_\parallel \le 0.2$~a.u. considered here, the DBOC and the field-induced changes in the electronic structure act, to a good approximation, as two independent contributions to the total energy rather than as strongly coupled effects.


This weak field dependence carries significant weight for high-resolution spectroscopy. Since $1$~cm$^{-1}\approx 30$~GHz, the DBOC-induced shifts found in this work correspond to frequency displacements ranging from several tens to a few hundred GHz, which fall well within the resolution limits of modern spectroscopic instruments. Such shifts are therefore highly relevant for modeling molecular lines in the atmospheres of magnetic white dwarfs. Field strengths of the magnitude considered here ($B_\parallel = 0.2$~a.u. $\approx 4.7\times 10^{8}$~G) are among the strongest observed for these compact objects~\cite{Kawka2020,Bagnulo2022,VeraRueda2024,Einramhof2026}, where accurate line positions are critical to determining both chemical abundances and magnetic field strengths.


Within $B_{\parallel}\le 0.2$~a.u. the field stretches the vibrational ladder by about $2.5\%$ for H$_2$, $2\%$ for HeH$^{+}$ and $7.5\%$ for LiH, that is, by hundreds of cm$^{-1}$, whereas the DBOC contributes at most $15$~cm$^{-1}$ and varies only weakly with the field. The two effects are therefore essentially additive, and the DBOC can be included as a nearly field-independent offset to the vibrational term values. Neglecting it would not mimic a field-dependent spectral shift, but would leave a systematic error of up to $15$~cm$^{-1}$ in the absolute line positions, which exceeds the $0.1$--$2.0$~cm$^{-1}$ accuracy currently attainable for molecular lines in the atmospheres of magnetic white dwarfs.

The field range explored here, $B_\parallel \le 0.2$~a.u. ($\approx 4.7\times 10^{4}$~T or $4.7\times 10^{8}$~G), is determined by the requirement that the singlet state remains the ground state. As shown in Ref.~\cite{10.1063/5.0269984}, above $B_\parallel \approx 0.2$~a.u., a repulsive triplet state drops below the singlet for both H$_2$ and LiH. Thus, our results are strictly restricted to fields below this crossover point. At the extreme fields characteristic of neutron stars ($10^{12}$--$10^{15}$~G or $10^{8}$--$10^{11}$~T), the electronic structure becomes far more strongly perturbed, and whether the decoupled behavior observed here persists remains an open question. A complementary development has recently been reported in Ref.~\cite{Yenugu2026}, where the nuclear Schr\"odinger equation for H$_2$ was solved on a three-dimensional grid for fields up to $3$~a.u. without restricting the molecular orientation. The present work approaches this challenge from the electronic structure side, utilizing correlated wavefunctions to quantify the adiabatic correction for heteronuclear and charged systems as well as for H$_2$. Bringing these two lines of development together represents the natural next step toward achieving spectroscopically accurate rovibrational data for highly magnetized astrophysical environments.

\bigskip

\section*{Data and Software Availability}
The data underlying this study are available in the published article and at
Zenodo (\url{https://doi.org/10.5281/zenodo.22792162}). The Python scripts used to
generate the integrals with ChronusQ, perform the UHF and UCISD calculations with
PySCF, evaluate the DBOC, and produce the figures, together with example input and
output files, are available at the same repository and at
\url{https://github.com/Physfock/dboc-magnetic-field}. ChronusQ and PySCF are
open-source software packages available from their developers.

\section*{Acknowledgments}
This work was supported by the Russian Science Foundation under grant 25-22-00643.

\bibliography{bibliography}
\end{document}